\documentclass[aps,prd,reprint,superscriptaddress,nofootinbib,floatfix]{revtex4-2}

\usepackage{amsmath,amssymb,amsfonts}
\usepackage{bm}
\usepackage{mathtools}
\usepackage{graphicx}
\usepackage[hidelinks]{hyperref}

\newtheorem{lemma}{Lemma}
\providecommand{\qed}{\hfill$\square$}
\newcommand{\Ric}{\mathrm{Ric}}
\newcommand{\EBY}{E_{\rm BY}}

\begin{document}

\title{Limits of Observer Choice for Brown--York Gravitational Energy}

\author{Hong-Xun Wang}
\affiliation{College of Physics, Jilin University, Changchun 130012, China}

\author{Samuel L. Braunstein}
\email{sam.braunstein@york.ac.uk}
\affiliation{Computer Science, University of York, York YO10 5GH, United Kingdom}

\author{Zhi-Wei Wang}
\email{zhiweiwang.phy@gmail.com}
\affiliation{College of Physics, Jilin University, Changchun 130012, China}

\date{\today}

\begin{abstract}
We ask how much freedom changing observers gives us to define gravitational energy, using the Euclidean-referenced Brown--York energy when the observers are the normals of a single spatial slice and must pass two tests at once, the vacuum small-sphere limit and the mass at infinity. Combining the Fan--Shi--Tam small-sphere expansion with the Gauss--Codazzi relations, we express the fifth-order vacuum coefficient through the electric and magnetic Weyl tensors and the extrinsic curvature of the slice. In spherical vacuum the slices with vanishing third-order term form a one-parameter scalar-flat family on which the normalized local curvature weight $W$ and the asymptotic energy obey the exact relation $W=\tfrac32[\EBY(\infty)/M]^2$, so that choosing a local normalization constrains the global result. For arbitrary spherical slices this becomes an exact radial balance law, the spherical equality case of the Geroch identity along inverse mean curvature flow, which ties the mass at infinity to accumulated local small-sphere contributions. Under rotation, no nondegenerate scalar-flat Schwarzschild slicing continues perturbatively into Kerr through second order in spin while retaining its own constant local curvature weight. Every such member is rigid at first order and fails an explicit monopole compatibility condition at second order. The rotational obstruction does not depend on which such weight is chosen and survives a second-order drift of the weight. Comparisons with the Brown--Lau--York foliation and with the Liu--Yau, Epp and Wang--Yau energies delimit the scope of the result.
\end{abstract}

\maketitle

\section{Introduction}
\label{sec:intro}

Energy cannot be localized unambiguously in a curved spacetime. General relativity provides a standard total mass for an isolated system, but no unique local density of gravitational energy, and the energy assigned to a finite boundary can depend on the observers and on the reference geometry used to define it \cite{Szabados}. A similar observer dependence is carried by generalized Komar charges and by the local surface gravity defined for a family of observers \cite{WB26,WB26H}. This dependence is not itself a defect. The question is whether observer freedom can pass independent tests. If changing observers repairs an energy's response to curvature in small vacuum regions, can the same observers retain the expected mass far away, and does that choice survive rotation?

We make this question concrete using the Euclidean-referenced Brown--York energy,
\begin{equation}
\EBY(S)=\frac{1}{8\pi}\oint_S(k_0-k)\,dA ,
\label{eq:BY}
\end{equation}
where $k$ is the mean curvature of the boundary $S$ in a spatial slice and $k_0$ is that of its isometric embedding in Euclidean space. The subtraction is held fixed, and we vary the hypersurface-orthogonal observer congruence by changing the slice, which also fixes the small geodesic spheres that probe the energy locally. Observer dependence of this kind is well known. Lau described boosts as changes of spanning slice \cite{Lau96}, Brown, Lau and York developed the associated boost laws \cite{BLY02}, Epp proposed a boost-invariant alternative \cite{Epp}, and covariant-Hamiltonian approaches have explored joint optimization of observer and reference choices \cite{Wu10,LCN11,WCLN12}. The Wang--Yau energy goes further and fixes the observer separately for each surface through an optimal isometric embedding into Minkowski space \cite{WY09,WY09b}. We ask instead what a single observer field, defined by one slice for all surfaces at once, can achieve when it must pass more than one test.

The local test is the vacuum small-sphere limit. A natural positive measure of the local Weyl curvature is the Bel--Robinson superenergy $T(u,u,u,u)=|E|^2+|H|^2$, with $E$ and $H$ the electric and magnetic Weyl tensors seen by an observer $u$. Small light-cone spheres yield this structure for the Hawking and Brown--York constructions \cite{HS,BLY99}, with coefficient $1/90$ for the latter when a light-cone reference is used, and the small-sphere limit of the Wang--Yau energy, with the observer fixed by its optimal embedding equation, is likewise governed by the Bel--Robinson tensor \cite{CWY18}. Its role for small-region gravitational energy has long been recognized \cite{Krishnasamy,KTW}. The coefficient is \emph{not} universal for geodesic spheres within spatial slices. On time-symmetric vacuum slices the spatial Hawking energy vanishes through fifth order, whereas the Brown--York energy retains a term proportional to $|E|^2$ \cite{FST}. We therefore use the light-cone coefficient as an explicit \emph{benchmark}, not as a required normalization, and pair it with the second test, the standard mass $M$ at infinity. Their compatibility is the issue.

Spherical vacuum symmetry makes this problem exactly solvable, since by Birkhoff's theorem its nonflat regions are locally Schwarzschild. We find an exact balance law that ties the local small-sphere response to the mass encoded by larger spheres, so a chosen local normalization has a definite cost at infinity. We then perturb toward Kerr and find that no \emph{nondegenerate} scalar-flat spherical member survives to second order in the spin while preserving its own constant local curvature weight. The freedom that permits local tuning in spherical symmetry becomes rigid under rotation, whatever constant weight is chosen.

The paper is organized as follows. Section~\ref{sec:functional} derives the vacuum congruence functional that expresses the fifth-order small-sphere coefficient through the Weyl tensors and the extrinsic curvature of the slice. Section~\ref{sec:spherical} solves the spherical problem on the scalar-flat family and obtains the exact relation between the local weight and the energy at infinity, together with the Komar energy of the slice observers. Section~\ref{sec:balance} extends the trade-off to arbitrary spherical slices as a radial balance law, relates it to the Geroch identity, and compares known observer choices. Section~\ref{sec:other} examines the Liu--Yau, Epp and Wang--Yau energies. Section~\ref{sec:kerr} contains the perturbative Kerr analysis, and Sec.~\ref{sec:conclusions} summarizes. Appendix~\ref{app:numbers} records exact verification values for the small-sphere law, Appendix~\ref{app:sources} gives the second-order Kerr sources and a quadrupole cross-check, and Appendix~\ref{app:code} describes the verification code.

We use geometric units. Throughout, $\epsilon$ denotes the radius of a geodesic sphere $S_\epsilon$ within a slice and $[\epsilon^n]F$ the coefficient of $\epsilon^n$ in the small-$\epsilon$ expansion of $F$. The areal radius of a spherical slice is $R$, the scalar curvature of a slice is written ${}^3R$, and $r$ is reserved for the Boyer--Lindquist radial coordinate of Sec.~\ref{sec:kerr}.

\section{Small-sphere expansion and the vacuum congruence functional}
\label{sec:functional}

\subsection{Fan--Shi--Tam expansion}

Let $S_\epsilon$ be a geodesic sphere of radius $\epsilon$ in a spatial slice $\Sigma$ with induced metric $\gamma_{ij}$ and Levi-Civita connection $D$. Fan, Shi and Tam proved the intrinsic expansion \cite{FST}
\begin{align}
\EBY(S_\epsilon)&=\frac{{}^3R}{12}\epsilon^3
+\Bigl[\frac1{60}|{}^3\Ric|^2-\frac{13}{1440}({}^3R)^2\nonumber\\
&\qquad+\frac1{120}\Delta{}^3R\Bigr]\epsilon^5+O(\epsilon^6),
\label{eq:FST}
\end{align}
where ${}^3\Ric$ is the Ricci tensor of $\gamma_{ij}$, $\Delta=D^kD_k$, and all curvature quantities are evaluated at the center of the sphere. The expansion involves only the intrinsic geometry of the slice. The observers enter through the slice itself. Its future-pointing unit normal $n$ defines the hypersurface-orthogonal congruence, and its extrinsic curvature $K_{ij}$ is the projection of $-\nabla_an_b$ onto $\Sigma$, with trace $K$.

In vacuum the Gauss--Codazzi relations give
\begin{equation}
{}^3R=K_{ij}K^{ij}-K^2,\qquad
{}^3\Ric_{ij}=E_{ij}-Q_{ij},
\label{eq:GC}
\end{equation}
with
\begin{equation}
Q_{ij}=KK_{ij}-K_{ik}K^k{}_j ,
\label{eq:Q}
\end{equation}
where $E_{ij}$ is the electric Weyl tensor seen by the slice normal. Equations~\eqref{eq:GC} and \eqref{eq:Q} express every term of Eq.~\eqref{eq:FST} through $E_{ij}$ and $K_{ij}$ except the derivative term $\Delta{}^3R$, which is the nontrivial step.

\subsection{Decomposition of the derivative term}

Let $T_{kij}=D_kK_{ij}$, with totally symmetric part $S_{kij}=T_{(kij)}$ and traces $v^{(1)}_i=D^kK_{ki}$ and $v^{(2)}_i=D_iK$. Under $SO(3)$ the tensor $T\in3\otimes\mathrm{Sym}^2$ splits into $S_{kij}$ and a part antisymmetric in its first two indices, which is equivalent to $\varepsilon_i{}^{kl}D_kK_{lj}$. By the Codazzi equation the symmetric part of the latter is, in vacuum and up to a sign fixed by convention, the magnetic Weyl tensor $H_{ij}$, while its antisymmetric part is dual to $v^{(1)}-v^{(2)}$. The irreducible decomposition therefore gives
\begin{equation}
|DK|^2=|S_{kij}|^2+\frac23|H|^2+\frac13|v^{(1)}-v^{(2)}|^2 ,
\label{eq:decompfull}
\end{equation}
with $|DK|^2\equiv D_kK_{ij}D^kK^{ij}$. The vacuum momentum constraint $D^kK_{ki}=D_iK$ sets $v^{(1)}=v^{(2)}$, so that
\begin{equation}
|DK|^2=|D_{(k}K_{ij)}|^2+\frac23|H|^2 .
\label{eq:decomp}
\end{equation}
Expanding the Laplacian of ${}^3R=K_{ij}K^{ij}-K^2$ gives $\Delta{}^3R=2K^{ij}\Delta K_{ij}+2|DK|^2-2K\Delta K-2|DK|^2_{\rm tr}$, with $|DK|^2_{\rm tr}\equiv D_kK\,D^kK$, and Eq.~\eqref{eq:decomp} then yields
\begin{equation}
\Delta{}^3R=2\mathcal D+\frac43|H|^2,
\label{eq:DeltaR}
\end{equation}
where
\begin{equation}
\mathcal D=K^{ij}\Delta K_{ij}-K\Delta K+|D_{(k}K_{ij)}|^2-|DK|_{\rm tr}^2 .
\label{eq:Ddef}
\end{equation}

\subsection{Congruence functional and the reduced condition}

Substituting Eqs.~\eqref{eq:GC} and \eqref{eq:DeltaR} into Eq.~\eqref{eq:FST} yields the general vacuum congruence functional
\begin{align}
{[\epsilon^5]}\EBY
&=\frac1{90}|H|^2+\frac1{60}|E-Q|^2+\frac1{60}\mathcal D \nonumber\\
&\quad-\frac{13}{1440}(K_{ij}K^{ij}-K^2)^2 .
\label{eq:functional}
\end{align}
The terms $\tfrac1{60}\mathcal D$ and $\tfrac1{90}|H|^2$ are not independent, since Eq.~\eqref{eq:DeltaR} fixes $\mathcal D$ in terms of $|H|^2$ wherever $\Delta{}^3R$ is known. On a slice with ${}^3R\equiv0$ they cancel exactly, leaving
\begin{equation}
{[\epsilon^5]}\EBY=\frac1{60}|{}^3\Ric|^2 .
\label{eq:scalarflat}
\end{equation}
At a point where the $\epsilon^3$ term vanishes, so that $K_{ij}K^{ij}=K^2$, the complete Bel--Robinson requirement $[\epsilon^5]\EBY=\tfrac1{90}(|E|^2+|H|^2)$ reduces to the single scalar condition
\begin{equation}
|E-Q|^2+\mathcal D=\frac23|E|^2.
\label{eq:weight}
\end{equation}
On a time-symmetric vacuum slice, $K_{ij}=0$ gives ${}^3\Ric_{ij}=E_{ij}$ and
\begin{equation}
\EBY(S_\epsilon)=\frac1{60}E_{ij}E^{ij}\,\epsilon^5+O(\epsilon^6),
\label{eq:timesym}
\end{equation}
which is the term proportional to $|E|^2$ mentioned in Sec.~\ref{sec:intro}. Exact checks of Eq.~\eqref{eq:FST} and of these reductions on explicit slices, including slices with $K_{ij}\neq0$ and a rotating example, are collected in Appendix~\ref{app:numbers}.

\section{The scalar-flat family in spherical vacuum}
\label{sec:spherical}

\subsection{Slices with vanishing third-order term}

By Birkhoff's theorem every nonflat spherically symmetric vacuum region is locally Schwarzschild with mass $M$, so the calculation below represents the full spherical-vacuum sector. A spherical slice with spacetime areal radius $R$ has the metric
\begin{equation}
\gamma=\frac{dR^2}{1-2m(R)/R}+R^2d\Omega^2 ,
\label{eq:massaspectmetric}
\end{equation}
which defines its mass aspect $m(R)$, and ${}^3R=4m'(R)/R^2$. A vanishing $\epsilon^3$ term at every point therefore forces $m(R)=m_0$ constant. These scalar-flat slicings are a known family \cite{OMR06,Akbar17,Bini15}; the new question is what the Bel--Robinson criterion selects within it. In Schwarzschild coordinates the member with mass aspect $m_0\le M$ is the graph $t=h_0(R)$ with
\begin{equation}
h_0'(R)=\mp\frac{\sqrt{2(M-m_0)}\,R}{(R-2M)\sqrt{R-2m_0}} ,
\label{eq:h0}
\end{equation}
where the upper sign gives the ingoing member and the lower sign its time reverse. The family contains the static slice $m_0=M$ and the flat Painlev\'e--Gullstrand slice $m_0=0$, and it continues to negative $m_0$.

For the ingoing member the extrinsic-curvature eigenvalues are
\begin{equation}
\kappa_R=-\frac12\sqrt{\frac{2(M-m_0)}{R^3}},\qquad
\kappa_\theta=\kappa_\phi=\sqrt{\frac{2(M-m_0)}{R^3}},
\label{eq:Kfamily}
\end{equation}
while the slice-normal Weyl curvature and intrinsic Ricci tensor are
\begin{equation}
E_{\hat a\hat b}=\frac{M}{R^3}\mathrm{diag}(1,1,-2),\qquad
{}^3\Ric_{\hat a\hat b}=\frac{m_0}{R^3}\mathrm{diag}(1,1,-2)
\label{eq:EandRic}
\end{equation}
in an orthonormal frame with the radial direction last. Hence $H=0$, $Q=(1-m_0/M)E$ and, by Eq.~\eqref{eq:DeltaR}, $\mathcal D=0$. The round symmetry sphere has mean curvature $k=2\sqrt{1-2m_0/R}/R$ in the slice and $k_0=2/R$ in Euclidean space, so its Euclidean Brown--York energy is exactly
\begin{equation}
\EBY(R)=R\!\left(1-\sqrt{1-\frac{2m_0}{R}}\right),
\qquad \EBY(\infty)=m_0 .
\label{eq:BYfamily}
\end{equation}

\subsection{Bel--Robinson selection}

The local condition now becomes algebraic. Equation~\eqref{eq:scalarflat} gives
\begin{equation}
{[\epsilon^5]}\EBY=\frac{m_0^2}{10R^6},
\qquad
\frac1{90}(|E|^2+|H|^2)=\frac{M^2}{15R^6},
\label{eq:weights}
\end{equation}
so the benchmark Bel--Robinson normalization holds at every exterior point if and only if
\begin{equation}
m_0=\pm\sqrt{\frac23}\,M .
\label{eq:roots}
\end{equation}
The positive member is a ``tempered-rain'' congruence between static observers and Painlev\'e--Gullstrand rain, and the negative member is a faster, intrinsically negative-mass member of the same family. By Birkhoff's theorem these are the only spherically symmetric congruence-wide repairs in spherical vacuum. The significance is not that the scalar-flat family is new, but that a local curvature criterion selects two exact observer congruences from it. Since $Q_{ij}=\lambda E_{ij}$ with $\lambda=1-m_0/M$ throughout the family, the selection reads $(1-\lambda)^2=2/3$ and can be stated covariantly in the initial data.

\subsection{Exact local--global relation}

The same closed form exposes an obstruction. We define the normalized local weight $W$ of a slice with vanishing $\epsilon^3$ term as the ratio of its $\epsilon^5$ coefficient to the benchmark. On the scalar-flat family this weight and the mass measured at infinity satisfy the exact relation
\begin{equation}
W\equiv\frac{{[\epsilon^5]}\EBY}{\tfrac1{90}(|E|^2+|H|^2)}
=\frac32\left(\frac{\EBY(\infty)}{M}\right)^2 .
\label{eq:localglobal}
\end{equation}
The benchmark thus forces $\EBY(\infty)=\pm\sqrt{2/3}\,M\neq M$, whereas the static slice $m_0=M$ has the standard mass but overweights the local electric sector by the factor $3/2$. Equation~\eqref{eq:localglobal} does not depend on the benchmark. If the normalized $\epsilon^5$ coefficient is required to take any nonnegative target value $W$, the family returns $\EBY(\infty)=\pm M\sqrt{2W/3}$, so every target other than the static value $W=3/2$ is paid for in the mass at infinity. None of this violates ADM invariance. For $m_0\neq M$ one has $K_{ij}=O(R^{-3/2})$, outside the standard asymptotic initial-data class, and the induced three-geometry itself has ADM mass $m_0$.

\subsection{Komar energy of the slice observers}

The same value appears as a Komar charge of the observers. For $\xi=Nn$, with $N$ the lapse and $n$ the unit normal of the slice, the generalized Komar energy of Ref.~\cite{WB26} on a two-sphere $S$ in the slice reduces, because $n^a\nabla_an_b=D_b\ln N$ and $n^b\nabla_an_b=0$, to
\begin{equation}
E_{\rm K}[Nn]=\frac1{4\pi}\oint_S s^iD_iN\,dA ,
\label{eq:komar}
\end{equation}
where $s$ is the outward unit normal of $S$ within the slice and the normalization is fixed so that the Killing vector $\partial_t$ gives $M$. The integrand $s^iD_iN=N\,a_is^i$, with $a_i=D_i\ln N$ the acceleration of the slice-normal observers, is a Komar surface density of the redshift-times-acceleration form that defines the Noether surface gravity of Ref.~\cite{WB26H}, evaluated here for the observers selected by the slice. On the scalar-flat member with mass aspect $m_0$ the lapse is $N=\sqrt{1-2m_0/R}$ on either branch, and $s^R=\sqrt{1-2m_0/R}$, so $s^iD_iN=m_0/R^2$ and
\begin{equation}
E_{\rm K}[Nn](S_R)=m_0
\label{eq:komarfamily}
\end{equation}
on every round sphere, whereas the Komar energy of $\partial_t$ equals $M$ on the same spheres. Equation~\eqref{eq:komarfamily} follows directly from the surface integral and equals both the spatial ADM mass $m_0$ and $\EBY(\infty)$. This agreement does not require a claim that the momentum-evolution flux of Ref.~\cite{WB26} vanishes. Coordinate stationarity under $\partial_t=Nn+\beta$, with nonzero radial shift $\beta$, does not imply vanishing normal evolution, and standard asymptotic ADM--Komar hypotheses need not apply to these slices. The two charges differ because their generators differ by the shift. The density $s^iD_iN$ has the redshifted-acceleration form of Ref.~\cite{WB26H}, but no dynamical-horizon temperature is inferred. The same observer parameter thus controls two desirable normalizations in incompatible ways.

\section{Radial balance law across scales}
\label{sec:balance}

\subsection{Balance law for arbitrary spherical slices}

The trade-off extends beyond the scalar-flat family and becomes an exact balance law across scales. For an arbitrary spherical slice, not necessarily scalar flat, let $\mathcal E(R)$ be the Euclidean Brown--York energy of the round symmetry sphere at areal radius $R$. Its mean curvature in the slice is $k=2\sqrt{1-2m(R)/R}/R$, while the Euclidean reference has $k_0=2/R$, so
\begin{align}
\mathcal E(R)&=R\left(1-\sqrt{1-\frac{2m(R)}{R}}\right),\nonumber\\
m(R)&=\mathcal E(R)-\frac{\mathcal E(R)^2}{2R}.
\label{eq:balanceinvert}
\end{align}
Let $p_R$ be any point on that symmetry sphere and define $c_3(R)\equiv[\epsilon^3]\EBY(S_\epsilon(p_R))$ for an infinitesimal geodesic sphere centered at $p_R$. Equation~\eqref{eq:FST} gives
\begin{equation}
c_3(R)=\frac{{}^3R(R)}{12}=\frac{m'(R)}{3R^2}.
\label{eq:c3}
\end{equation}
Differentiating Eq.~\eqref{eq:balanceinvert} therefore gives the exact identity
\begin{equation}
\frac{d}{dR}\left[\mathcal E(R)-\frac{\mathcal E(R)^2}{2R}\right]
=3R^2c_3(R).
\label{eq:radialbalance}
\end{equation}
If $m(R)\to M$ at infinity, integration gives
\begin{equation}
M-\mathcal E(R)+\frac{\mathcal E(R)^2}{2R}
=3\int_R^\infty \rho^2c_3(\rho)\,d\rho .
\label{eq:sumrule}
\end{equation}
The difference between the mass aspect at radius $R$ and its asymptotic value is thus exactly the accumulated $\epsilon^3$ defect outside that radius.

\subsection{Hawking mass and the Geroch identity}

The combination $\mathcal E-\mathcal E^2/(2R)$ is the \emph{Riemannian} Hawking mass of $S_R$, built from the mean curvature $k$ within the chosen slice. With $A=4\pi R^2$,
\begin{align}
m_H(S_R)&=\sqrt{\frac{A}{16\pi}}\left(1-\frac1{16\pi}\oint k^2dA\right)\nonumber\\
&=\frac R2\left[1-\left(1-\frac{2m(R)}R\right)\right]=m(R).
\label{eq:hawking}
\end{align}
This is not the spacetime Hawking (Misner--Sharp) mass of the same sphere. The latter replaces $k^2$ by $k^2-\tau^2$, where $\tau\equiv\sigma^{AB}K_{AB}$ is the trace of the slice extrinsic curvature over the sphere with induced metric $\sigma_{AB}$, and it equals $M$ on every Schwarzschild symmetry sphere independently of the slicing.

For a smooth inverse mean curvature flow of spheres with $k>0$ in a three-manifold, parametrized by the logarithm of the area $A$, the Geroch evolution identity reads \cite{Geroch73,HI01}
\begin{equation}
\frac{dm_H}{d\ln A}=\frac{\sqrt{A}}{(16\pi)^{3/2}}\oint\Bigl({}^3R+2|\nabla\log k|^2+|\mathring k_{AB}|^2\Bigr)dA ,
\label{eq:geroch}
\end{equation}
where $\nabla$ is the gradient on the flow surface and $\mathring k_{AB}$ its trace-free second fundamental form. The round spheres of a spherical slice are an inverse mean curvature flow with $d\ln A=2\,dR/R$, on which $k$ is constant and $\mathring k_{AB}=0$, so Eq.~\eqref{eq:geroch} becomes $dm_H/dR=R^2\,{}^3R/4=3R^2c_3$, which is Eq.~\eqref{eq:radialbalance}. Since ${}^3R=12[\epsilon^3]\EBY$ pointwise by Eq.~\eqref{eq:FST}, the identity balances the Hawking mass against the small-sphere defect. The gradient and shear terms are nonnegative. Along any smooth inverse mean curvature flow of spheres with $k>0$ in a vacuum slice, the Hawking mass gained is therefore bounded below by the accumulated small-sphere defect, and Eq.~\eqref{eq:sumrule} is the case of equality. This is a comparison identity rather than a monotonicity statement, because ${}^3R=K_{ij}K^{ij}-K^2$ need not be nonnegative on a general vacuum slice. General Brown--York variation formulas are known \cite{MST10}.

\begin{figure*}[t]
\centering
\includegraphics[width=0.96\textwidth]{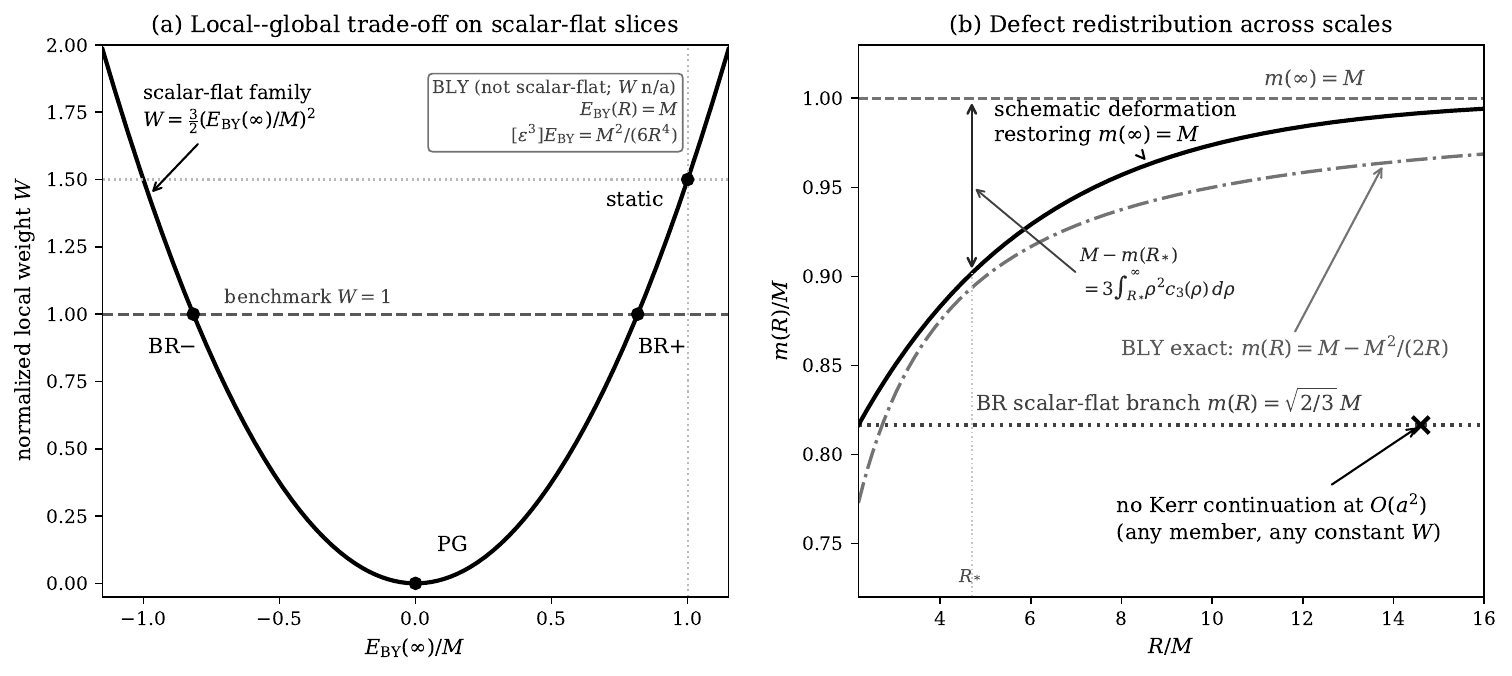}
\caption{Observer freedom redistributes rather than removes the Brown--York mismatch. (a) On the scalar-flat family the normalized local weight and asymptotic Brown--York mass obey $W=\tfrac32[\EBY(\infty)/M]^2$. The light-cone Bel--Robinson benchmark $W=1$ selects the rigid branches $\EBY(\infty)=\pm\sqrt{2/3}M$, whereas the standard mass selects the static value $W=3/2$. The Brown--Lau--York (BLY) foliation lies off this curve because it has $[\epsilon^3]\EBY=M^2/(6R^4)\neq0$. (b) Schematic radial balance law. The positive Bel--Robinson branch has the frozen mass aspect $m(R)=\sqrt{2/3}M$, and any spherical deformation restoring $m(\infty)=M$ must accumulate the defect of Eq.~\eqref{eq:sumrule}; the exact BLY curve is shown for comparison. The cross marks the separate perturbative Kerr obstruction for nondegenerate members at their own constant weights and is not a trajectory in this plot.}
\label{fig:observerlandscape}
\end{figure*}

Equation~\eqref{eq:sumrule} makes the tension sharp. On the positive Bel--Robinson branch $c_3=0$ wherever the member is retained, so any spherical deformation that ultimately restores $m(\infty)=M$ requires the net defect
\begin{equation}
3\int \rho^2c_3(\rho)\,d\rho=M\left(1-\sqrt{\frac23}\right)
\label{eq:netdefect}
\end{equation}
over the region in which the mass aspect is changed, as illustrated in Fig.~\ref{fig:observerlandscape}.

\subsection{Known observer choices}

Known observer choices make the trade-off concrete. Brown, Lau and York exhibited a boosted Schwarzschild foliation on which the referenced Brown--York energy equals $M$ on every round sphere \cite{BLY02}. Its radial metric coefficient is $(1-M/R)^{-2}$, so it has $m(R)=M-M^2/(2R)$ and hence ${}^3R=2M^2/R^4\neq0$. These observers reproduce the finite-radius and global mass but carry a nonzero leading small-sphere term already at order $\epsilon^3$, the converse of the observers in Eq.~\eqref{eq:roots}. Equation~\eqref{eq:radialbalance} gives $c_3=M^2/(6R^4)$ directly from $\mathcal E(R)=M$, and Eq.~\eqref{eq:sumrule} integrates this defect to $M^2/(2R)=M-m(R)$ exactly. Painlev\'e--Gullstrand rain is more extreme still, since its spatial metric is flat and Eq.~\eqref{eq:BY} vanishes identically. Observer freedom is therefore large, but in spherical symmetry it redistributes the mismatch rather than moving all desiderata independently. Table~\ref{tab:tradeoff} summarizes the comparison.

\begin{table}[t]
\caption{Four observer families for Schwarzschild Brown--York energy, with $W$ the normalized $\epsilon^5$ coefficient of Eq.~\eqref{eq:localglobal} when the $\epsilon^3$ term vanishes. The BLY row denotes the boosted foliation of Ref.~\cite{BLY02}; its nonzero $\epsilon^3$ term makes $W$ inapplicable. The negative Bel--Robinson branch has the same $W$ and $\EBY(\infty)/M=-\sqrt{2/3}$.}
\label{tab:tradeoff}
\begin{ruledtabular}
\begin{tabular}{lccc}
observers & $[\epsilon^3]\EBY$ & $W$ & $\EBY(\infty)/M$\\
\colrule
static & $0$ & $3/2$ & $1$\\
Bel--Robinson $+$ & $0$ & $1$ & $\sqrt{2/3}$\\
PG rain & $0$ & $0$ & $0$\\
BLY $E=M$ & $M^2/(6R^4)$ & -- & $1$\\
\end{tabular}
\end{ruledtabular}
\end{table}

\subsection{Pointwise versus congruence-wide conditions}

The congruence-wide qualifier is essential. At an isolated radius $R_p$, a vanishing $\epsilon^3$ term requires only $m'(R_p)=0$. Writing $m_p=m(R_p)$, $m_2=m''(R_p)$ and $m_3=m'''(R_p)$, the $\epsilon^5$ coefficient at such a point is
\begin{align}
{[\epsilon^5]}\EBY\big|_p&=\frac{1}{30R_p^6}\Bigl[3m_p^2+m_2R_p^2(5m_p-2R_p)\nonumber\\
&\qquad\qquad+m_3R_p^3(R_p-2m_p)\Bigr].
\label{eq:onepoint}
\end{align}
The free derivatives $m_2$ and $m_3$ can be chosen so that this coefficient takes the benchmark value while $m(R)\to M$ at infinity. Our no-go concerns a congruence that passes the benchmark \emph{wherever its observers are} and retains the standard mass at infinity, which is what a preferred observer field defined throughout a region requires. The balance law shows the escape route, since a pointwise repair fits into a globally correct slice only if $c_3$ becomes nonzero elsewhere, with its weighted integral fixed by Eq.~\eqref{eq:sumrule}.

\section{Other energy prescriptions}
\label{sec:other}

\subsection{Liu--Yau energy}

Changing the energy prescription also relocates rather than removes the tension. We use the boost-invariant Liu--Yau energy \cite{LY03} as a diagnostic comparison. With the same Euclidean reference embedding as Brown--York, it is
\begin{equation}
E_{\rm LY}=\EBY+\frac1{8\pi}\oint\left(k-\sqrt{k^2-\tau^2}\right)dA ,
\label{eq:LYdef}
\end{equation}
with $\tau$ the trace of the slice extrinsic curvature over the surface, as in Sec.~\ref{sec:balance}. For a small geodesic sphere with outward unit normal $\nu^i$ in the slice, $k=2/\epsilon+O(\epsilon)$ and $\tau=K-K_{ij}\nu^i\nu^j+O(\epsilon)$. Hence $k-\sqrt{k^2-\tau^2}=\tau^2/(2k)+O(\epsilon^3)$. Using the angular average
\begin{equation}
\langle \nu_i\nu_j\nu_k\nu_l\rangle=\frac{\delta_{ij}\delta_{kl}+\delta_{ik}\delta_{jl}+\delta_{il}\delta_{jk}}{15}
\label{eq:average}
\end{equation}
and adding the Brown--York term $(K_{ij}K^{ij}-K^2)\epsilon^3/12$ gives
\begin{equation}
E_{\rm LY}(S_\epsilon)=\frac1{10}|\varsigma|^2\epsilon^3+O(\epsilon^5),
\label{eq:LY}
\end{equation}
where $\varsigma_{ij}=K_{ij}-\tfrac13K\gamma_{ij}$. The trace of $K_{ij}$ drops out exactly, and only the shear $\varsigma_{ij}$ of the congruence remains.

For the round spheres $S_R$ of a spherical slice, $\sqrt{k^2-\tau^2}$ is the norm of the spacetime mean curvature vector of $S_R$ and does not depend on the slice. In Schwarzschild it equals $(2/R)\sqrt{1-2M/R}$, so $E_{\rm LY}(S_R)=R(1-\sqrt{1-2M/R})$ for every member of the scalar-flat family and at every radius, not only in the large-sphere limit. On the same family $|\varsigma|^2=3(M-m_0)/R^3$, so the leading Liu--Yau term of small geodesic spheres is positive on every member except the static one, even though the finite-radius and large-sphere Liu--Yau values are those of the static slice for the entire family. This is the small-sphere counterpart of the known non-rigidity of Kijowski--Liu--Yau mass \cite{OMST04}. The issue is therefore how the observer and reference prescriptions are coupled.

\subsection{Epp and Wang--Yau energies}

The slice independence of the round-sphere value rests only on the norm of the spacetime mean curvature vector of $S_R$ and on a round reference. It therefore also holds for Epp's invariant quasi-local energy \cite{Epp}, which is likewise built from this boost-invariant norm. For a round symmetry sphere the normal-bundle curvature vanishes, so the round embedding in a constant-time slice of Minkowski space satisfies Epp's matching condition for the reference, and with this reference every member of the scalar-flat family again assigns the static value to its round spheres. We have not evaluated the small-sphere expansion of Epp's energy on these slices and make no claim about it.

The Wang--Yau energy \cite{WY09,WY09b} goes further and couples the two prescriptions explicitly, fixing the observer for each surface by solving an optimal embedding equation into Minkowski space. Its small-sphere limit along light cones is governed by the Bel--Robinson tensor in vacuum, which completes the verification of its consistency with the classical limits \cite{CWY18}. Because the optimal observer is chosen separately for each surface, it need not coincide with the unit normal of any single spatial slice containing both the small geodesic spheres and the large round spheres. The obstruction of this paper concerns one slice-defined observer field used at all scales, and it therefore does not apply to the Wang--Yau construction. The two results are complementary. They indicate that consistency across scales is obtained when the observer and the reference are fixed jointly for each surface, and is not available from a single slicing with a fixed Euclidean reference.

\section{Rotation and the perturbative Kerr obstruction}
\label{sec:kerr}

We now ask whether the spherical freedom survives rotation, with Kerr the first controlled test. Throughout this section and Appendix~\ref{app:sources} we set $M=1$, so that $a$, $r$ and $m_0$ are measured in units of $M$, and use Boyer--Lindquist coordinates $(t,r,\theta,\phi)$ with $x=\cos\theta$. At $a=0$ the coordinate $r$ coincides with the areal radius $R$ of Secs.~\ref{sec:spherical} and \ref{sec:balance}. All computations in this section are exact and are reproduced by the code of Ref.~\cite{zenodo}.

\subsection{Reduced conditions and perturbative setup}

On any slice with ${}^3R=0$, Eq.~\eqref{eq:DeltaR} gives $\mathcal D=-2|H|^2/3$, so Eq.~\eqref{eq:weight} is equivalent to
\begin{equation}
\mathcal S\equiv|{}^3\Ric|^2-\frac23|E|^2-\frac23|H|^2=0 .
\label{eq:S}
\end{equation}
Away from symmetry the pair ${}^3R=0$ and $\mathcal S=0$ constrains one graph function by two scalar equations, yet in spherical symmetry the system collapses to the single equation $m'=0$ and one algebraic condition on $m_0$. The question is whether this compatibility survives the first physical breaking of spherical symmetry.

The background is the scalar-flat member of Eq.~\eqref{eq:h0}, which for $M=1$ has
\begin{equation}
h_0'(r)=\frac{c\,r}{(r-2)\sqrt{r-2m_0}},\qquad c=\sqrt{2(1-m_0)} ,
\label{eq:h0kerr}
\end{equation}
on the outgoing branch $h_0'>0$. The ingoing member differs by time reversal and gives identical scalars. We consider a completely general graph slicing of Kerr near this background,
\begin{equation}
t=h_0(r)+a\,h_1+a^2h_2+o(a^2),
\label{eq:hexpansion}
\end{equation}
with $h_1$ and $h_2$ general functions on the slice. The discrete symmetry $(a,\phi)\mapsto(-a,-\phi)$ maps the background slice to itself and makes the scalar sources on the undeformed slice even in $a$. It does not force an individual solution branch to be even, so $h_1$ must be retained. At order $a$ there is no source, and $h_1$ obeys the homogeneous linearized equations. Lemma~\ref{lem:kernel} below shows that $h_1$ is then a constant, that is, a time translation. Since a constant $h_1$ is a Killing time translation of Kerr for every $a$, all terms quadratic in $h_1$ and all cross terms between $h_1$ and the $O(a)$ metric correction drop out of the $O(a^2)$ equations exactly. These equations are therefore $\mathcal L[h_2]+(\text{second-order source})=0$, with the sources evaluated on the undeformed slice $t=h_0(r)$.

The sources follow from the full four-dimensional Kerr metric on the undeformed slice and satisfy ${}^3R=O(a^2)$, $\mathcal S=(6m_0^2-4)/r^6+O(a^2)$ and $H=O(a)$, with
\begin{equation}
|H|^2=\frac{18a^2\,[r(1+2x^2)-2m_0(1-x^2)]}{r^9}+O(a^4).
\label{eq:Hsq}
\end{equation}
The $O(a^2)$ sources are axisymmetric and equatorially even, and their $\ell=4$ projections vanish identically (Appendix~\ref{app:sources}). Since the linearized operators are diagonal in $\ell$ and all unsourced harmonics of $h_2$ vanish by Lemma~\ref{lem:kernel}, it suffices to take $h_2=\chi_0(r)+\chi_2(r)P_2(x)$.

\subsection{Linearized operators and first-order rigidity}

To avoid confusion with the magnetic Weyl tensor $H_{ij}$, we denote the radial coefficient of a perturbation by $\chi(r)$. For
$h=h_0(r)+\mu\,\chi(r)Y_{\ell m}$, with $L=\ell(\ell+1)$, the terms of the two scalars linear in $\mu$ are
$\delta{}^3R=\mathcal L^{(\ell)}_R[\chi]\,Y_{\ell m}$ and $\delta\mathcal S=\mathcal L^{(\ell)}_{\mathcal S}[\chi]\,Y_{\ell m}$
with, for arbitrary $c$ and hence arbitrary $m_0=1-c^2/2$,
\begin{align}
\mathcal L^{(\ell)}_R[\chi]
&=\frac{c\sqrt{r-2m_0}}{r^4}\Bigl[L\chi-2(r+4m_0)\chi'\nonumber\\
&\hspace{5.2em}-4r(r-2m_0)\chi''\Bigr],\nonumber\\
\mathcal L^{(\ell)}_{\mathcal S}[\chi]
&=-\frac{2m_0c\sqrt{r-2m_0}}{r^7}
\Bigl[-\tfrac52L\chi+(5r-16m_0)\chi'\nonumber\\
&\hspace{5.2em}-2r(r-2m_0)\chi''\Bigr].
\label{eq:ops}
\end{align}
The operators were derived in two independent ways, by the ADM deformation of the background slice along the vector $\mu\chi Y_{\ell m}\partial_t$, decomposed into its parts normal and tangential to the slice, and directly from the induced metric and extrinsic curvature of a general graph slice $t=h(r,x)$. Both give Eq.~\eqref{eq:ops}. The terms linear in $\mu$ that are proportional to derivatives of $Y_{\ell m}$, and the $x$ dependence of the coefficients of $Y_{\ell m}$, vanish identically once the Legendre equation is used, as rotational invariance requires. Since the background is spherically symmetric the radial operators do not depend on $m$, so nonaxisymmetric harmonics are covered. A shift $\delta m_0$ along the family gives $\delta{}^3R=0$ and $\delta\mathcal S=12m_0\delta m_0/r^6$.

The superenergy $B\equiv|E|^2+|H|^2$ does not contribute to $\mathcal L^{(\ell)}_{\mathcal S}$. Under the perturbation the slice normal tilts by a spatial vector $\delta n$ and the slice is displaced along $\partial_t$. The displacement does not change $B$ because $\partial_t$ is Killing, while the tilt changes $B=T_{abcd}n^an^bn^cn^d$ by $4T_{abcd}n^an^bn^c\delta n^d$, which is proportional to the super-Poynting vector $\varepsilon^{ijk}E_{jl}H^l{}_k$ \cite{Senovilla00,MB98} and vanishes because $H=0$ on every spherical slice of Schwarzschild. A direct computation confirms that the coefficients of $\chi$, $\chi'$ and $\chi''$ in $\delta B$ vanish identically in $m_0$ for $\ell=0,1,2$. Thus $\mathcal L^{(\ell)}_{\mathcal S}$ is the linearization of $|{}^3\Ric|^2$ alone, and Eq.~\eqref{eq:ops} holds for every member of the family, independently of the weight attached to $B$ in Sec.~\ref{sec:weights}.

Eliminating the common second derivative gives the combination
\begin{align}
-\frac{r^3}{m_0}\mathcal L_{\mathcal S}^{(\ell)}[\chi]-\mathcal L_R^{(\ell)}[\chi]
&=\frac{6c\sqrt{r-2m_0}}{r^4}\nonumber\\
&\quad\times\left[2(r-2m_0)\chi'-L\chi\right],
\label{eq:firstorder}
\end{align}
which is of first order.

\begin{lemma}\label{lem:kernel}
For $m_0c\ne0$ and every $\ell\ge1$ the homogeneous system
$\mathcal L^{(\ell)}_R[\chi]=\mathcal L^{(\ell)}_{\mathcal S}[\chi]=0$ has only the
solution $\chi\equiv0$. For $\ell=0$ its solutions are the constants.
\end{lemma}

\emph{Proof.} By Eq.~\eqref{eq:firstorder}, $\chi=Cf_L$ with $f_L=(r-2m_0)^{L/2}$, and
\begin{equation}
\mathcal L^{(\ell)}_R[f_L]
=-\frac{L\,c\,[(L-2)r+6m_0]}{r^4\sqrt{r-2m_0}}\,f_L ,
\label{eq:kernelproof}
\end{equation}
which is not identically zero for $L\ge2$ when $m_0c\ne0$. Hence
$C=0$. For $L=0$ the first-order equation reads
$2(r-2m_0)\chi'=0$. \qed

The lemma is algebraic and uses no boundary or regularity condition. It is a rigidity statement. At $a=0$ both Bel--Robinson-selected congruences, and every member with $m_0c\neq0$, are rigid against every first-order slicing deformation that preserves ${}^3R=0$ and the local weight, nonaxisymmetric ones included, whatever the benchmark value. The only freedom at $\ell=0$ is a constant time translation. The $\delta m_0$ direction,
$\chi'=\partial_{m_0}h_0'=-r/[c\,(r-2m_0)^{3/2}]$, spans the kernel of $\mathcal L^{(0)}_R$
alone, whose general solution is $\chi'\propto r(r-2m_0)^{-3/2}$,
and Eq.~\eqref{eq:ops} gives $\mathcal L^{(0)}_{\mathcal S}[\chi]=12m_0/r^6$ on it, reproducing $\delta\mathcal S=12m_0\delta m_0/r^6$. Every harmonic that could generate a compensating $O(a^2)$ source is thus eliminated before the Kerr source is considered, so the second-order monopole test below is decisive for the full second-order perturbative problem.

\subsection{Second-order monopole obstruction}

At order $a^2$ the no-go already appears in the axisymmetric monopole. Let $S^{(\ell)}_R$ and $S^{(\ell)}_{\mathcal S}$ be the projections onto $P_\ell(x)$ of the $O(a^2)$ parts of ${}^3R$ and $\mathcal S$ on the undeformed slice, so that the second-order equations read $\mathcal L^{(\ell)}_R[\chi_\ell]+S^{(\ell)}_R=0$ and $\mathcal L^{(\ell)}_{\mathcal S}[\chi_\ell]+S^{(\ell)}_{\mathcal S}=0$. With sources, Eq.~\eqref{eq:firstorder} gives
\begin{align}
2(r-2m_0)\chi_\ell'-L\chi_\ell&=\eta_\ell ,\nonumber\\
\eta_\ell&\equiv\frac{r^4\bigl[S^{(\ell)}_R+(r^3/m_0)S^{(\ell)}_{\mathcal S}\bigr]}{6c\sqrt{r-2m_0}} .
\label{eq:eta}
\end{align}
For $\ell=0$ this fixes $\chi_0'=\eta_0/[2(r-2m_0)]$, and what remains is the compatibility residual
\begin{equation}
\mathcal R_0(r;m_0)\equiv\Bigl\{\mathcal L^{(0)}_R[\chi_0]+S^{(0)}_R\Bigr\}_{\chi_0'=\eta_0/[2(r-2m_0)]} .
\label{eq:residualdef}
\end{equation}
Inserting the closed-form monopole sources of Eqs.~\eqref{eq:SR0} and \eqref{eq:SS0} gives, for arbitrary $m_0$,
\begin{equation}
\mathcal R_0(r;m_0)=-\frac{4\,[2(1+m_0^2)r+3m_0^2-8m_0]}{m_0r^6} .
\label{eq:residual}
\end{equation}
The poles of the individual sources at $r=2$ cancel in this combination. At $m_0=\pm\sqrt{2/3}$ the residual reduces to
\begin{equation}
\mathcal R_0=-\frac{4}{3m_0r^6}(10r+6-24m_0),
\label{eq:residualBR}
\end{equation}
whose only zero, $r=(12m_0-3)/5$, lies below the throat $r=2m_0$ of the positive branch and at negative $r$ for the negative branch. The second-order equations are therefore incompatible on every nonempty open shell, including the weak-field region, and neither Bel--Robinson-selected congruence admits a continuation into Kerr to second order in the spin that preserves both conditions. The quadrupole sector provides an independent nonzero remainder (Appendix~\ref{app:sources}), but the obstruction does not depend on it.

\subsection{Arbitrary constant and drifting weights}
\label{sec:weights}

The obstruction does not depend on the benchmark either. Let $\mathcal S_W\equiv|{}^3\Ric|^2-\tfrac{2W}{3}B$ for a constant $W$, so that $\mathcal S_1=\mathcal S$. Replacing the target weight $1$ by $W$ leaves the operators of Eq.~\eqref{eq:ops}, the combination \eqref{eq:firstorder} and Lemma~\ref{lem:kernel} unchanged, because the linearization of $B$ vanishes, and it modifies only the $O(a^2)$ source. The $O(a^2)$ projections of $B$ on the undeformed slice are, for arbitrary $m_0$,
\begin{equation}
B^{(0)}=\frac{6(3r-8m_0)}{r^9},\qquad B^{(2)}=-\frac{12(3r-4m_0)}{r^9},
\label{eq:Bsrc}
\end{equation}
with vanishing $\ell\ge4$ projections. On the axis $x=1$ the sum $B^{(0)}+B^{(2)}=-18/r^8$ is independent of $m_0$, as it must be, since there every spherical normal lies in the principal plane of the type D field and $B=6|\Psi_2|^2=6/(r^2+a^2)^3$ for such observers. The monopole source of $\mathcal S_W$ is $S^{(0)}_{\mathcal S}-\tfrac23(W-1)B^{(0)}$, and the elimination through Eq.~\eqref{eq:firstorder} gives
\begin{align}
\mathcal R_0^{(W)}(r;m_0)&=\mathcal R_0(r;m_0)-\frac{8(W-1)(r-4m_0)}{m_0r^6}\nonumber\\
&=-\frac{4\,[2(W+m_0^2)r+3m_0^2-8Wm_0]}{m_0r^6},
\label{eq:residualW}
\end{align}
which reduces to Eq.~\eqref{eq:residual} at $W=1$. By Eq.~\eqref{eq:localglobal} a member satisfies $\mathcal S_W=0$ at $a=0$ only for its own weight $W=\tfrac32m_0^2$, and there
\begin{equation}
\mathcal R_0^{(3m_0^2/2)}(r;m_0)=-\frac{4m_0\,(5r+3-12m_0)}{r^6},
\label{eq:residualown}
\end{equation}
a nonzero rational function for every member with $m_0c\neq0$. Its only zero, $r=(12m_0-3)/5$, lies below the throat $r=2m_0$ for $0<m_0<1$ and at negative $r$ for $m_0<0$. Hence no nondegenerate scalar-flat spherical member satisfying its own constant weight at $a=0$ admits a continuation into Kerr to second order in the spin that retains that weight.

As an unrestricted algebraic equation, the numerator of Eq.~\eqref{eq:residualW} does vanish identically at the formal pair $(m_0,W)=(-3/8,-9/64)$. This is not an admissible background for the continuation, since the zeroth-order relation instead fixes $W=3m_0^2/2=27/128$ at $m_0=-3/8$. No nondegenerate member satisfying its own zeroth-order local weight can therefore eliminate the residual, regardless of which constant weight was selected.

Nor does the obstruction rely on keeping the target weight exactly fixed as rotation is introduced. A drift $W(a)=W_0+w_2a^2+O(a^4)$ contributes, through the background value $B=6/r^6$, only the additional monopole source
\begin{equation}
\delta S^{(0)}_{\mathcal S}=-\frac{4w_2}{r^6},\qquad \delta S^{(0)}_R=0.
\label{eq:driftsource}
\end{equation}
By Eq.~\eqref{eq:eta} this shifts the monopole solution by
\begin{equation}
\delta\chi_0'=-\frac{w_2r}{3m_0c(r-2m_0)^{3/2}}
=\frac{w_2}{3m_0}\,\partial_{m_0}h_0' .
\label{eq:driftchi}
\end{equation}
The extra deformation $\delta h_2=(w_2/3m_0)\,\partial_{m_0}h_0$ is exactly tangent to the scalar-flat Schwarzschild family. Since $\mathcal L_R^{(0)}[\partial_{m_0}h_0]=0$ and $\mathcal L_{\mathcal S}^{(0)}[\partial_{m_0}h_0]=12m_0/r^6$, it cancels the source of Eq.~\eqref{eq:driftsource} without changing the compatibility remainder, and Eq.~\eqref{eq:residualW} is unchanged for arbitrary constant $w_2$. Allowing the target normalization to drift at $O(a^2)$ merely moves the solution along the spherical scalar-flat family and cannot cure the Kerr obstruction.

\subsection{Scope and relation to earlier results}

The two degenerate parameter values $m_0=0$ (flat rain) and $m_0=1$ (static) lie outside this analysis, which assumes $m_0c\neq0$. At $m_0=0$ the prefactor of $\mathcal L^{(\ell)}_{\mathcal S}$ vanishes but $\mathcal L^{(\ell)}_R$ need not, whereas at $m_0=1$ the factor $c=0$ makes both operators of Eq.~\eqref{eq:ops} vanish. For $m_0=0$ the requirement $W=0$ on scalar-flat slices forces $|{}^3\Ric|^2=0$, hence a flat three-geometry, and the spatially flat Kerr no-go of Ref.~\cite{DFLMO19} is a separate result with its own assumptions. The result of Garat and Price \cite{GP00} assumes a static Schwarzschild limiting slice and does not directly apply to the rain endpoint. The present perturbative argument does not settle either endpoint.

The obstruction is distinct from these earlier no-go results for spatially and conformally flat Kerr slicings \cite{GP00,DFLMO19}, and Brown--York energy in Kerr and other constrained Kerr slicings are themselves well studied \cite{Martinez94,Schinkel14}. Our statement is local in the space of slicings and spacetimes. It excludes second-order continuation of every nondegenerate, locally weighted Schwarzschild member on any open exterior shell, but not a disconnected, strongly nonspherical solution with no Schwarzschild limit. It needs no asymptotic boundary conditions or horizon regularity and is already visible in the local field equations.

\section{Conclusions}
\label{sec:conclusions}

The freedom to choose observers does not allow the local and large-scale Brown--York energies to be tuned independently. In spherical vacuum the same observer choice fixes the small-sphere curvature weight and the asymptotic mass, and the radial balance law records exactly how any change in the spatial mass aspect must be accompanied by an accumulated leading small-sphere contribution. The Brown--Lau--York foliation gives the expected mass on every round sphere but incurs a lower-order local term, while the observers selected by the light-cone Bel--Robinson benchmark remove that term and assign a different mass at infinity, which is also the generalized Komar energy of those observers \cite{WB26}. This is neither a violation of ADM invariance nor a claim that the benchmark is compulsory. Rotation sharpens the restriction, since every nondegenerate scalar-flat member satisfying its own constant local weight is rigid at first order and fails the Kerr compatibility test at second order, even with a second-order drift of the weight.

The result has definite limits. It concerns a single hypersurface-orthogonal observer field with the Euclidean reference held fixed. The Kerr analysis is perturbative to second order in the spin about nondegenerate members, and it leaves open the endpoints $m_0=0$ and $m_0=M$ as well as strongly nonspherical solutions with no Schwarzschild limit. Within these limits, a construction that aims to reconcile local and global gravitational-energy tests must address the observer field and the reference subtraction together, as the Wang--Yau energy does surface by surface \cite{WY09,CWY18}. \emph{Changing the observers changes the energy, but does not give independent control of its answers at every scale.}

\begin{acknowledgments}
Z.-W.W. used Claude Opus 5.5 (Anthropic, 2026) to check some analytical calculations in the manuscript. The authors take full responsibility for the content of this manuscript.
\end{acknowledgments}

\section*{Data availability}
The code that verifies the analytic results and the quoted values of this paper is openly available from Ref.~\cite{zenodo}.

\appendix

\section{Exact verification values for the small-sphere law}
\label{app:numbers}

All values in this appendix are exact, and $M=1$. They record the invariant inputs and the resulting small-sphere coefficients for a set of test slices. The code of Ref.~\cite{zenodo} obtains each coefficient in two independent ways, from exact metric jets inserted into Eq.~\eqref{eq:FST} and directly from geodesic spheres built by the exponential map and embedded in Euclidean space as surfaces of revolution. The $S^3$ and $H^3$ entries are also compared with closed forms.

\emph{Static Schwarzschild slice.} On this time-symmetric slice the point at areal radius $R=\tfrac83$ has
\begin{equation}
{[\epsilon^5]}\EBY=\tfrac{729}{2621440},\qquad |E|^2=\tfrac{2187}{131072},
\end{equation}
and the point at $R=\tfrac{25}8$ has
\begin{equation}
{[\epsilon^5]}\EBY=\tfrac{131072}{1220703125},\qquad
|E|^2=\tfrac{1572864}{244140625}.
\end{equation}
In both cases $[\epsilon^5]\EBY/|E|^2=\tfrac1{60}$ exactly and $[\epsilon^1]\EBY=\cdots=[\epsilon^4]\EBY=0$. On time-symmetric vacuum slices ${}^3\Ric_{ij}=E_{ij}$ is trace free and divergence free, every admissible scalar at this order reduces to $|E|^2$, and the two evaluations, which have different values of $M^2/R^6$, overdetermine the coefficient. This confirms Eq.~\eqref{eq:timesym}, the $K_{ij}=0$ sector of the congruence functional \eqref{eq:functional}.

\emph{Ingoing Kerr--Schild slice of Schwarzschild.} At the point $p$ on the axis with $z_0=\tfrac{32}9$, $z$ denoting the Kerr--Schild height along the symmetry axis,
\begin{equation}
\EBY=\tfrac{2187}{1280000}\,\epsilon^3
+\tfrac{28603748943}{409600000000000}\,\epsilon^5+O(\epsilon^6),
\label{eq:KSschw}
\end{equation}
with, from independent jets, ${}^3R(p)=\tfrac{6561}{320000}$ and
$[\epsilon^3]\EBY/{}^3R(p)=\tfrac1{12}$ exactly. The invariant triple is
\begin{align}
&\Bigl({}^3R,\;|{}^3\Ric|^2,\;\Delta{}^3R\Bigr)(p)\nonumber\\
&\qquad=\Bigl(\tfrac{6561}{320000},\;
\tfrac{221610897}{204800000000},\;
\tfrac{2134798497}{320000000000}\Bigr).
\label{eq:KStriple}
\end{align}

\emph{Kerr--Schild slice of Kerr at order $a^2$.} At the same point on the rotation axis, $\Psi_2=-M/(z_0-ia)^3$ has boost weight zero, so the
slice-normal Weyl data are
$|E|^2=(6M^2/z_0^6)(1-12a^2/z_0^2)+O(a^4)$ and
$|H|^2=54M^2a^2/z_0^8+O(a^4)$. The invariants from exact degree-4
jets are
\begin{align}
{}^3R(p;a)&=\tfrac{6561}{320000}
-\tfrac{1241977617}{131072000000}\,a^2,\nonumber\\
|{}^3\Ric|^2(p;a)&=\tfrac{221610897}{204800000000}
-\tfrac{26248900391217}{20971520000000000}\,a^2,\nonumber\\
\Delta{}^3R(p;a)&=\tfrac{2134798497}{320000000000}
-\tfrac{660018714011973}{104857600000000000}\,a^2,
\label{eq:kerrinv}
\end{align}
up to $O(a^4)$, and inserting them into Eq.~\eqref{eq:FST} gives
\begin{align}
{[\epsilon^5]}\EBY(p;a)&=\tfrac{28603748943}{409600000000000}\nonumber\\
&\quad-\tfrac{1171159181183979}{16777216000000000000}\,a^2+O(a^4).
\label{eq:kerrr5}
\end{align}
The exact relative
spin response of the $\epsilon^5$ coefficient is
$[a^2\epsilon^5]/[a^0\epsilon^5]=-\tfrac{734580873}{734863360}\approx-0.99962$, against a
pure electric response $-12/z_0^2=-\tfrac{243}{256}\approx-0.94922$. Under
Eq.~\eqref{eq:FST} the response is the sum of the contributions \mbox{$\approx-0.299$},
$+0.050$ and $-0.751$ of the $|{}^3\Ric|^2$, $({}^3R)^2$ and $\Delta{}^3R$ terms. It is
dominated by the $\Delta{}^3R$ term, that is, by the extrinsic-curvature sector of the
congruence functional, and differs from the pure electric response by \mbox{$\approx-0.0504$}.

\emph{Further checks of the small-sphere law.} Flat space in a nontrivial
chart gives
\begin{equation}
(|{}^3\Ric|^2,({}^3R)^2,\Delta{}^3R)=(0,0,0),\qquad [\epsilon^5]\EBY=0.
\end{equation}
For
$\gamma=\delta+(x_2^2+x_3^2)\,e_1e_1$,
\begin{align}
(|{}^3\Ric|^2,({}^3R)^2,\Delta{}^3R)&=(6,16,24),\nonumber\\
[\epsilon^5]\EBY&=\tfrac7{45},
\end{align}
while for $\gamma=(1+x_1+x_1^2/2)^2\delta$,
\begin{align}
(|{}^3\Ric|^2,({}^3R)^2,\Delta{}^3R)&=(2,4,-24),\nonumber\\
[\epsilon^5]\EBY&=-\tfrac{73}{360}.
\end{align}
The round $S^3$ and hyperbolic $H^3$ of unit radius both give $(12,36,0)$ and
$[\epsilon^5]\EBY=-\tfrac18$, in agreement with the closed forms
$\sin\epsilon(1-\cos\epsilon)$ and $\sinh\epsilon(1-\cosh\epsilon)$.

\section{Second-order sources and the quadrupole cross-check}
\label{app:sources}

This appendix lists the $O(a^2)$ sources of Sec.~\ref{sec:kerr}, with $M=1$. For arbitrary $m_0$ the monopole projections are
\begin{align}
S^{(0)}_R
&=-\frac{4}{r^6(r-2)^3}\Bigl[
2(m_0-1)r^4\nonumber\\
&\qquad-(4m_0^2-7m_0+6)r^3\nonumber\\
&\qquad-(8m_0^2-22m_0-4)r^2\nonumber\\
&\qquad-4m_0(4m_0+5)r+8m_0(4m_0-1)\Bigr],
\label{eq:SR0}
\end{align}
\begin{align}
S^{(0)}_{\mathcal S}
&=\frac{4}{r^9(r-2)^3}\Bigl[
(2m_0^2-5m_0-3)r^4\nonumber\\
&\qquad-(8m_0^3-23m_0^2-2m_0-18)r^3\nonumber\\
&\qquad-(8m_0^3-10m_0^2+32m_0+36)r^2\nonumber\\
&\qquad-(16m_0^3+68m_0^2-96m_0-24)r\nonumber\\
&\qquad+64m_0^3+8m_0^2-64m_0\Bigr],
\label{eq:SS0}
\end{align}
and the quadrupole projections are
\begin{align}
S^{(2)}_R
&=-\frac{4\bigl[(1-m_0)r+m_0(4m_0-1)\bigr]}{r^6},
\label{eq:SR2}\\
S^{(2)}_{\mathcal S}
&=-\frac{4}{r^9}\bigl[(7m_0^2+8m_0-6)r\nonumber\\
&\qquad\quad-8m_0^3-m_0^2+8m_0\bigr].
\label{eq:SS2}
\end{align}
The $\ell=4$ projections vanish identically. The poles at $r=2$ are artifacts of
expanding $\Delta^{-1}=(r^2-2r+a^2)^{-1}$ in $a$ in Boyer--Lindquist
coordinates, and they cancel in the monopole residual \eqref{eq:residual}. These closed forms hold identically in $m_0$, as verified with $c$ kept symbolic in Ref.~\cite{zenodo}. Equations~\eqref{eq:Bsrc}--\eqref{eq:residualown} are likewise verified identically in $m_0$ by the direct four-dimensional computation of Ref.~\cite{zenodo}.

The $\ell=2$ sector is a cross-check and is not a necessary part of the obstruction. For $\ell=2$, substituting $\chi_2'$ and $\chi_2''$ from Eq.~\eqref{eq:eta} into the equation $\mathcal L^{(2)}_R[\chi_2]+S^{(2)}_R=0$ gives an algebraic relation $p(r)\chi_2+q(r)=0$ with $p\ne0$, and the remainder is the residual of the first-order equation on $\chi_2=-q/p$. Independent differential elimination of the $\ell=0$ and
$\ell=2$ systems gives nonzero remainders. The monopole remainder is exactly
Eq.~\eqref{eq:residual}. At the rational member $m_0=41/50$ it becomes
$-2(8362r-11357)/(1025r^6)$, reproducing the factor found by the independent
elimination. The quadrupole remainder is likewise nonzero, and at $m_0=41/50$
it contains the factor
$42142500r^3+103911425r^2-175501443r-229093404$. At the two Bel--Robinson members every unsourced $\ell\ge1$ harmonic vanishes by Lemma~\ref{lem:kernel}.

\section{Verification code}
\label{app:code}

\begin{sloppypar}
The exact-arithmetic code of Ref.~\cite{zenodo} reproduces every analytic statement and quoted value of this paper and generates Fig.~\ref{fig:observerlandscape}. Running \texttt{run\_all.py} executes all checks. Sections~\ref{sec:functional}--\ref{sec:other} and Appendix~\ref{app:numbers} are checked by the scripts \texttt{c1\_decomp.py}, \texttt{c4\_family.py}, \texttt{c5b\_onepoint.py}, \texttt{c14\_komar\_check.py}, \texttt{c3\_liuyau.py} and \texttt{c2\_small\_sphere.py}. Section~\ref{sec:kerr} and Appendix~\ref{app:sources} are checked by \texttt{c12\_kernel\_ops.py} and \texttt{c12\_kernel.py}, by the chain \texttt{c10\_op\_r.py}, \texttt{c8\_source.py} and \texttt{c11\_elim\_r.py} at $m_0=41/50$, by the independent graph-slice pipeline \texttt{c13\_general\_weight.py} with $c$ kept symbolic, and by \texttt{c15\_weights.py} for the formal pair and the drifting weight of Sec.~\ref{sec:weights}.
\end{sloppypar}

\end{document}